\documentclass[pre,floatfix]{revtex4-1} 
\usepackage[pdfpagemode=UseNone,pdfstartview=FitH,colorlinks=true]{hyperref}
\usepackage{enumitem}
\setlistdepth{9}
\usepackage{graphicx}
\usepackage{epstopdf}

\def\<{{\langle}}
\def\>{{\rangle}}

\begin{document}
\title{Motor function in interpolar microtubules during metaphase}

\author{J. M. Deutsch}
\email{josh@ucsc.edu}
\author{Ian P. Lewis}
\email{iplewis@ucsc.edu}
\affiliation{Department of Physics, University of California, Santa Cruz CA 95064}

\begin{abstract}
We analyze experimental observations of microtubules undergoing
small fluctuations about a ``balance point" when mixed in solution
of two different kinesin motor proteins, KLP61F and Ncd. It has
been proposed that the microtubule movement is due to stochastic
variations in the densities of the two species of motor proteins.
We test this hypothesis here by showing how it maps onto a
one-dimensional random walk in a random environment. Our estimate
of the amplitude of the fluctuations agrees with experimental
observations. We point out that there is an initial transient in
the position of the microtubule where it will typically move of
order its own length. We compare the physics of this gliding assay
to a recent theory of the role of antagonistic motors on restricting
interpolar microtubule sliding of a cell's mitotic spindle during
prometaphase. It is concluded that randomly positioned antagonistic
motors can restrict relative movement of microtubules, however they
do so imperfectly. A variation in motor
concentrations is also analyzed and shown to lead to greater control
of spindle length.
\end{abstract}
\maketitle
 
\section{Introduction}

During mitosis, pole spacing is regulated by a system of interpolar
microtubules. It has been proposed that the interpolar microtubules can be moved
in two directions by opposing motors, but the details of such a
proposed system are not yet well known.
The interpolar microtubules are likely bundled and moved by two
families of kinesin motor proteins; kinesin-5 and kinesin-14.
Experiments with \textit{Drosophila melanogaster} suggest that a
kinesin-5 motor protein, KLP61F, plays a large role in creating the
spindle during prometaphase \cite{heck}. It has also been shown
that kinesin-5 forms cross-bridges between interpolar microtubules
in the centralspindlin \cite{sharp}. Further experiments suggest
the same motor drives the separation of the poles during metaphase
and anaphase \cite{brust,li}. In vitro experiments show that KLP61F
slides antiparallel microtubules apart on motility assays, where
motor proteins are bound to glass slides and move microtubules that are
added to the solution \cite{li}.

All of the above results show that kinesin-5 plays an important role
in controlling the spindle spacing. Being a tetramer with both
dimers at the N-terminus, the motor can walk toward the plus ends
of  two antiparallel microtubules, thus forcing the poles apart.
\\

The kinesin-5 are antagonized by the kinesin-14, which walk toward
the  minus end of the microtubules. In vitro experiments show a
kinesin-14, Ncd, is capable of bundling microtubules and driving
an inward sliding of the interpolar microtubules \cite{sharp}. With
one motor able to separate the poles, and one able to bring them
closer, it seems possible that the two motors are responsible for
maintaining spindle spacing and moving the poles apart. The net
force exerted by the two motor species could govern the direction
and rate of pole movement.  

Recently, work has been done in trying to understand how outward
microtubule sliding generated by the kinesin-5 and inward sliding
generated by the kinesin-14 could result in the stable, steady-state
spindle spacing during prometaphase. A balance of forces could
result in a stationary spindle, but it is unclear how the ``collective
antagonism" could occur \cite{li}. In the following section, we will
discuss one group's proposed solution to the problem.

\subsection{Experimental Work}

Experiments with in vitro motility assays were performed to see if
KLP61F and Ncd could interact to control the speed and polarity of
microtubules motility and whether the antagonism between the motors
could stall microtubule sliding enough to produce the stable
steady-state spindle spacing observed during prometaphase \cite{li}.
Before combining both motors in an assay, each motor was observed
moving microtubules in motility assays as expected. KLP61F moved
microtubules at $0.04 \mu m/s$ with the minus ends leading and Ncd
moved microtubules at $0.1 \mu m/s$ with the plus ends leading
\cite{li}. Further experiments also showed that KLP61F alone, Ncd
alone, and mixtures of the two motors bundled microtubules under
conditions with physiological ATP concentrations \cite{li}.  

So see how the two species of motors would interact, different molar
ratios of KLP61F and Ncd were mixed and microtubule motility was
measured. A balance point at a mole fraction of 0.7 Ncd was found
where microtubules displayed a mean velocity of approximately zero
\cite{li}. For greater mole fractions of Ncd, the mean velocity was
plus end directed. Conversely, for smaller mole fractions of Ncd, the mean
velocity was minus end directed, as shown in Fig. 5(a)
of Ref. \cite{li}. The slope of the lines fit to the two sides
of the balance point in this figure suggests that KLP61F
is a strong, slow motor that is not slowed down easily by the weak,
fast Ncd motor, which in turn is slowed down easily by KLP61F
\cite{li}. At the balance point, the microtubules where observed
to display oscillatory
motion between KLP61F and Ncd directed movement with intermediate
rates of roughly 0.02 $\mu$m/s, as shown in Fig. 5(b)
of Ref. \cite{li}.  

The authors in Ref. \cite{li} suggest that KLP61F and Ncd motors
could act synchronously to antagonize one another. However, being
an inherently stochastic process, it is hard to see how motor power
stroking could become synchronized. In later work~\cite{civelekoglu}
a fully stochastic model with many parameters was devised and tested
numerically. Ref. \cite{li} had suggested that
the microtubules could be gliding on a spatially varying landscape,
with varying densities of KLP61F and Ncd motors \cite{li}. Periods
of directional movement would be due to the patches in the environment
where one motor is dominant. It is possible the microtubule finds
a ``valley" in the landscape where it oscillates between patches of
motors that move it back towards the balance point. It is this
theory that we will attempt to model in the following section.

We show that the phenomenon is quite general and independent of
the details in the parameters. 
If the system is rescaled to be dimensionless in length and time, we find that
the behavior is only controlled by one parameter; the effective ``temperature" of
the system.
A detailed understanding of the motors will 
only change this effective temperature and nothing else, since scaling laws for
spatio-temporal fluctuations are universal.
A study from this perspective also elucidates
other aspects of this system, such as the nature of initial transients
in motion of the microtubules in these assays before they reach a quasi-steady state.
These transients have interesting implications, as we show that they also should occur
for interpolar microtubules during metaphase.

\section{Physical analysis of antagonistic motor assay}

\subsection{Average force-velocity dependence of antagonistic motors}

\begin{figure}[htb]
\begin{center}
\includegraphics[width=4.0in, angle=0]{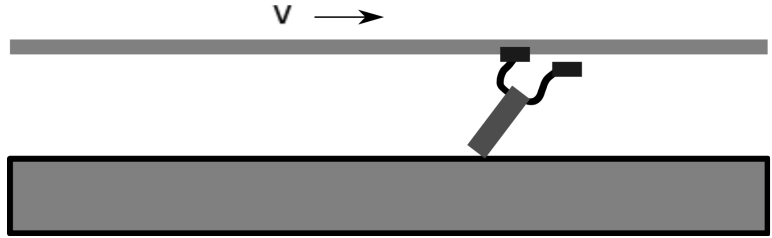}
\caption
{ 
A single motor with the lower end anchored against a glass plate, with its heads
binding and unbinding with a microtubule that is being moved at a constant
velocity $v$ relative to the plate.
}
\label{fig:SingleMotor}
\end{center}
\end{figure}

We first consider the problem of a single molecular motor, such as kinesin,
with the tail tethered to a substrate such as a glass plate while the heads can freely
interact with a long microtubule, as shown in Fig. \ref{fig:SingleMotor}. 
The microtubule is being moved along a single
dimension at uniform velocity $v$ that is parallel to the glass. The heads
bind and unbind with the microtubule, applying an average net force $f$, that 
will depend on $v$. The averaging is being done over time, and we considering
the limit where the time interval goes to infinity.

Now consider a collection of $N$ identical motors that all interact with the
same microtubule but are sufficiently distant from each other that they can
be considered independent. Then the average force acting on the microtubule
due to these motors is $N f$. 

\begin{figure}[htb]
\begin{center}
\includegraphics[width=4.0in, angle=0]{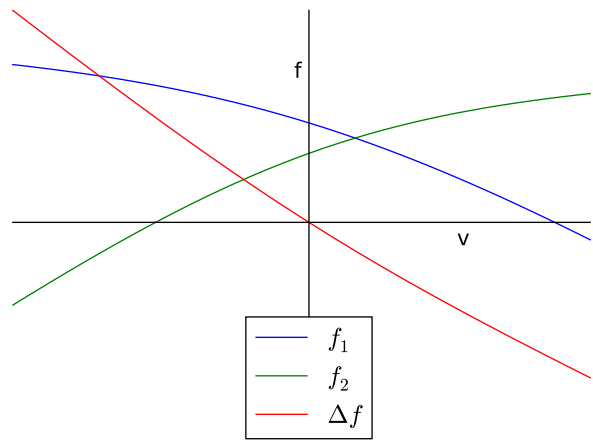}
\caption
{ 
The average force versus velocity curves, $f_1$ and $f_2$,for two species of motors, $1$ and $2$,
that act to antagonize each other. The weighted difference between these two curves, is $\Delta f$,
with relative concentrations chosen so that they are at the balance point,
so that $\Delta f=0$ for $v=0$.
}
\label{fig:fvsv}
\end{center}
\end{figure}

The above analysis is easily extended to the case of two separate species of
antagonistic motors, labeled $1$ and $2$, with average force versus velocity curves $f_1(v)$ and $f_2(v)$
respectively, as shown in Fig. \ref{fig:fvsv}.  If the number of motors of each 
kind is $N_1$ and $N_2$, then
the time averaged force is $N_1 f_1 - N_2 f_2$, where we have adopted
a sign convention so that the net force is a difference, rather than a sum.
If we choose the ratio $N_1/N_2 = f_2(0)/f_1(0)$, then this net average force
vanishes at $v=0$. This is the ``balance point" where there is no net average
force acting on the microtubule. The difference between $f_1$ and $f_2$, $\Delta f(v)$, 
weighted in this way is shown in Fig. \ref{fig:fvsv}.

Note that at small velocities at the balance point, Fig. \ref{fig:fvsv}, the net effect of these
motors, due to the linear relationship between force and velocity in this
regime, is, on average, to give linear drag. This linearity breaks down at high
velocities, but, as we will see below, we are interested in the low velocity
regime.  In this case
the net force $f_n$ is proportional to $\Delta f$ so that 
\begin{equation}
f_n = -\gamma v,
\end{equation}
where $\gamma$ is the drag coefficient.

Note that if we are not interested in small velocities, there are
other phenomena that can take place that can potentially invalidate
this analysis. For large enough velocities, the assumption of a
fixed force versus velocity curve can sometimes
fail~\cite{ProstJulicherPRL,GeurinProstMartinJoanny}. Some molecular
motors at an individual level, are capable of being in two internal
states which can be metastable.  These two states have different
characteristics, and on average, apply their net force in opposite
directions. The relative prevalence of the two states depend on the
microtubule velocity. If many such motors are simultaneously attached
to a microtubule, and allowed to pull it freely, this can lead to
the microtubule moving in one direction for a long time and then
occassionally, due to the collective fluctuation of all the motors,
reversing its direction of motion.

For the problem under discussion here, the two motors are known to
be antagonistic and are assumed to be near the balance point, so
that the average velocity is very small, and we are therefore
justified in taking a time average over very long times, which
will then give rise to a single $f$ versus $v$ curve. Therefore we believe that
the above behavior is not relevant to our particular case, although
it is possible that it could become so, for mixtures of other kinds of
motors, and therefore deserves further study. 

The above analysis is important for two reasons. First, it relates precisely the
force versus velocity curve for assays containing a motor mixture at
arbitrary concentration to the behavior of single motor assays. 
Empirical data on single motor assays can be obtained and then used to
understand average properties of these mixed systems. Second, it shows
that at the balance point there is only one parameter that need be
characterized in order to understand observations; the drag coefficient
$\gamma$. 

Now that we have understood the behavior of the average force,
we shall turn our attention to the effects of spatio-temporal fluctuations on the motion
of microtubules. As we shall see, these fluctuations are crucial to
understanding the system's behavior.

\subsection{The effect of spatiotemporal fluctuations}

We will consider a system at the balance point, where the concentration
of the two antagonistic motors has been adjusted as mentioned above, so that
the average force on a microtubule is zero. However even at this point, the
time averaged force acting on a mictrotubule will depend on its location. This
is because the motors are positioned randomly, so that there are fluctuations
in the net force. If on average, the number of motors acting on a microtubule
is $N$, then we expect this fluctuating time averaged force to have an
amplitude that varies proportional
to $\sqrt{N}$. This time averaged force will vary slowly as a function of
position. If the microtubule is moved only slightly, most of the same motors
will still act upon it, meaning that the force will be highly correlated
with its original value. The microtubule has to move its entire length before
this time averaged force becomes completely independent of its initial value.

Aside from this time averaged force that is position dependent, we can consider
a fixed position and ask how the force varies with time. There will be a
substantial variation in the force as a function of time due to the random
binding and unbinding of motor ends to the microtubule. Because these events
are uncorrelated between motors, the amplitude of these fluctuation will also
vary proportional to $\sqrt{N}$.

The above considerations imply that there are two parts to the force exerted
on the microtubule, a spatially varying component $F(x)$, and a temporal
component $n(t)$. For a fixed microtubule, the total force is the sum of
these two terms. The statistical properties of $n$ and $F$ are 
independent of each other because, for a long microtubule, $n$ is
the sum of many independent components and therefore its amplitude
is independent of position

If we now consider a microtubule that is no longer fixed in position, there
is a further force due to the drag, as discussed above.

With these three physical effects included, we are now in a position
to model this problem more precisely by characterizing the statistical
properties of $n(t)$ and $F(x)$, as we now discuss.

\subsection{Model as a Random Walk in Random Environment}

Both in a cell and in the experiments, the two motor proteins and
microtubules are mixed together in a solution. To model this simply
and in one dimension, we imagine a railroad track with motor proteins
randomly placed at every tie. This creates a random environment. A
rigid microtubule of length $L$, is placed on the tracks, and the motors that lie
underneath randomly exert a force on the microtubule. With both
species of motors randomly exerting forces on the microtubule in
opposing directions, the microtubule undergoes random movement on the track
given by
\begin{equation}
\gamma {\dot x} = F(x) + n(t),
\label{eqn:rwre}
\end{equation} 
where $F(x)$ is the random static force, $n(t)$ is the time fluctuating force, and $\gamma$ is the drag coefficient,
as discussed above. The left hand side contains $v = \dot x$, which is the
velocity of the center of mass of the microtubule.
Eq. \ref{eqn:rwre} is generally known as a Random Walk in a Random Environment \cite{rwre}. 
The difference between this and previous work lies in the correlations
in $F(x)$, which as noted above, is correlated over the length of a microtubule.

Below, we will analyze this equation as follows:
The random force $n(t)$ gives an effective temperature for this system.
By calculating the statistics of the force the motors exert on the
microtubule, this temperature can be determined. By calculating the statistics
of $F(x)$, we will know how potential correlations behave for length scales 
much less than $L$.  We can thereby estimate how far a microtubule will
move, on average, before being stopped by a potential barrier. Using
this, we can make estimates about the oscillatory behavior seen in
antagonistic gliding assays.

\subsection{Determining Forces Exerted on the Microtubule}

One railroad tie will by occupied by either a KLP61F or Ncd motor.
If we measure the force the motor exerts over a time much longer
than that motor's cycle, we will see an average force
\begin{equation}
\<f\>_{t}= f_{KLP},
\end{equation}
for one KLP61F motor, or
\begin{equation}
\<f\>_{t}=f_{NCD},
\end{equation}
for one Ncd motor, where $f_{NCD}<0$. Because the motor exerts a
peak force for some time and then exerts no force, the average force
is given by
\begin{equation}
f_{KLP}=f_{k}p_{k}
\end{equation}
for KLP61F, and
\begin{equation}
f_{NCD}=f_{n}p_{n},
\end{equation}
for Ncd, where $f_{k}$ and $f_{n}$ are the peak forces exerted by
the motors, and $f_{n}<0$. $p_{k}$ and $p_{n}$ are the probabilities
of the motors exerting a force on the microtubule. For simplicity,
we will set $p_{k}=p_{n}=p$.  
\\

The average force exerted by a single motor along the track is
determined by the concentrations of the motor species. For a
concentration $k$ of KLP61F, the force from one motor site averaged
over time and space is given by
\begin{equation}
\<\<f\>_{t}\>_{x} = kf_{k}p + (1-k)f_{n}p.
\label{eqn:staticf}
\end{equation}
For an average net force $\<F\>=0$, as seen experimentally, the variance of the static force is given by
\begin{equation}
\<\<f\>^{2}_{t}\>_{x} = k(f_{k}p)^{2}+(1-k)(f_{n}p)^{2}=p^{2}[kf_{k}^{2}+(1-k)f_{n}^{2}].
\label{eqn:staticvar}
\end{equation}.
\\

At each motor site, however, the average force is non-zero. Therefore
the variance of the time fluctuating force, $n(t)$, is given by
\begin{equation}
\<f^{2}\>-\<f\>^{2} = kf^{2}_{k}p(1-p)+(1-k)f^{2}_{n}p(1-p)=p(1-p)[kf_{k}^{2}+(1-k)f_{n}^{2}].
\label{eqn:fluctvar}
\end{equation}

\subsection{Behavior of the Potential for Distances $\ll L$}

Over distances greater than the length of the microtubule, $L$, the
potential will look like a random walk, but to determine if the microtubule
will fluctuate about a mean position, as seen experimentally, we must
look at the potential at scales $\ll L$.  

Because $\< F(x)\> = 0$,
we expect there to be many zeros for $F$ and the dynamics with $n(t)=0$ are such
that the microtubule will move downhill in potential to arrive at such points.
For finite $n(t)$, the microtubule will still on average move towards lower
points in potential, but will fluctuate around local potential minima.
Therefore we will look at the fluctuations of a microtubule after
it has moved into a position, $x^*$, where the $F(x^*)=0$.
This would represent a local minimum of the potential.  The question
we are addressing here is: are the fluctuations due to $n(t)$ small enough
to confine the microtubule to a certain region? In this section, for simplicity
we chose our coordinate system so that $x^* = 0$. 
Because the force, and therefore the potential, is finite everywhere,
and the statistics of the force are translationally invariant, the
microtubule cannot be localized to any one region and is expected
to eventually move arbitrarily far from an initial point.
However we will see that while this is true, the
time scale for this happening becomes extremely large, so in practice 
an experiment will observe confinement of the microtubule to particular
region. We will estimate the size of the region that would be explored
under normal experimental conditions.

If we move the microtubule a distance, $m$, we will see a difference
in the net force exerted on the microtubule. While most of the
microtubule is still being moved by the same motors, a length $m$
of it will be moved by new motors. Therefore, the potential will
be changed by some amount proportional to a factor of $m$.  

The net force exerted along the length of the microtubule is given by
\begin{equation}
F(m) = \sum_{i=\frac{-L}{2}+m}^{\frac{L}{2}+m}\eta_{i},
\label{eqn:fnet}
\end{equation} 
where $\eta_{i}$ is the force exerted at motor site $i$. Therefore
the difference in the net force is
\begin{equation}
F(m)-F(0) = \sum_{i=\frac{-L}{2}+m}^{\frac{L}{2}+m}\eta_{i}-\sum_{i=\frac{-L}{2}}^{\frac{L}{2}}\eta_{i}.
\label{eqn:pot1}
\end{equation}

Eq. \ref{eqn:pot1} simplifies because of cancellations on the right hand side, giving
\begin{equation}
F(m)-F(0)=\sum_{j=0}^{m}\phi_{j},
\label{eqn:pot2}
\end{equation}
where $\phi$ has been introduced to simplify the right hand side
of Eq. \ref{eqn:pot1}. Note that all $\phi$'s used below are
independent and that $\<\phi^{2}\>=2\<\eta_{i}^{2}\>$. As discussed above,
we are interested in fluctuations about a potential minimum so that
$F(0) = 0$.

The potential difference is given by
\begin{equation}
V(x)=-\int^{x}_{0}F(x')dx'.
\label{eqn:pot3}
\end{equation}
Turning Eq. \ref{eqn:pot3} into a Riemann Sum, with segments $\Delta$ equal to the spacing between motors, gives
\begin{equation}
V(m)=-\Delta\sum_{i=0}^{m}f(x_{i})=\Delta\sum_{i=0}^{m}\sum_{j=0}^{j}\phi_{j}=\Delta\sum_{j=0}^{m}j\phi'_{j}.
\label{eqn:pot4}
\end{equation}

To see how the potential scales with $m$, we look at the average
potential difference squared, $\<(V(m)-V(0))^{2}\>$. To simplify,
we will calculate the fluctuations about the minimum, so that
$V(0)=0$. Therefore,
\begin{equation}
\<(V(m)-V(0))^{2}\>=\<V(m)^{2}\>=\Delta^{2}\sum_{j=0}^{m}\sum_{i=0}^{m}ij\<\phi'_{i}\phi'_{j}\>.
\label{eqn:pot5}
\end{equation}

The correlation function $\<\phi'_{i}\phi'_{j}\>$ can be determined
from the correlation function for individual motors,
$\<\eta_{i}\eta_{j}\>$. Since the motors are regularly spaced, the
correlation function is given by,
\begin{equation}
\<\eta_{i}\eta_{j}\>=c\delta_{ij},
\label{eqn:cor1}
\end{equation}
because the motors are correlated at distances equal to the motor
spacing, and uncorrelated otherwise. The constant c is equal to the
variance of the static force, given by Eq. \ref{eqn:staticvar}. Eq.
\ref{eqn:cor1} becomes \begin{equation}
\<\eta_{i}\eta_{j}\>=p^{2}[kf_{k}^{2}+(1-k)f_{n}^{2}]\delta_{ij}.
\label{eq:c}
\end{equation} \\\

Since
$\<\phi^{2}\>=\<(\eta_{1}+\eta_{2})^{2}\>=\<\eta^{2}_{1}\>+\<\eta^{2}_{2}\>$,
the variance of $\phi$ is twice the variance of $\eta$, so that,
\begin{equation} 
\<\phi_{i}\phi_{j}\>=2c\delta_{ij}.  
\label{eqn:cor2}
\end{equation}

Inserting Eq. \ref{eqn:cor2} into Eq. \ref{eqn:pot5} and considering large $m$ yields
\begin{equation}
\<V(m)^{2}\>=2c\Delta^{2}\sum_{i=0}^{m}i^{2}=2c\Delta^{2}\int_{o}^{m}x^{2}dx=\frac{2}{3}c\Delta^{2}m^{3}.
\label{eqn:potfluct}
\end{equation}
Therefore, the potential fluctuations for distances $m\ll L$ are proportional to $m^{3}$. To calculate how far on average a microtubule will move from a minimum potential, the effective temperature must be determined.

\subsection{Determining the Effective Temperature}

Determining the effective temperature of the system will tell us
the energy of the system and determine how high the potential
barriers must be to keep the microtubule trapped in a potential
well. We can do this following the standard argument used to show
the Fluctuation Dissipation theorem~\cite{SethnaBookFluctDiss}. Determining the diffusion coefficient, $D$, will allow us to
make use of the Einstein Relation,
\begin{equation}
D=\frac{k_{B}T}{\gamma}.
\label{eqn:einstein}
\end{equation}

The force on the microtubule is given by $f(x,t)=\gamma v$, therefore,
\begin{equation}
\int_{0}^{t}f(t)dt=\gamma(x(t)-x(0)).
\end{equation}
Therefore,
\begin{equation}
\<\gamma^{2}(x(t)-x(0))^{2}\>=\<(\int_{0}^{t}f(t')dt')^2\>=\int_{0}^{t}\int_{0}^{t}\<f(t')f(t")\>dtdt'.
\label{eqn:diff1}
\end{equation}

The correlation function, $\<f(t)f(t')\>$, can be approximated as a delta function, such that
\begin{equation}
\<f(t)f(t')\>=2c_{0}\tau_{d}\delta(t-t'),
\label{eqn:corf}
\end{equation}
where $c_{0}=p(1-p)[kf_{k}^{2}+(1-k)f_{n}^{2}]$, the variance of
the time fluctuating force given by Eq. \ref{eqn:fluctvar} and
$\tau_{d}$ is the decay time of a motor. Substituting Eq. \ref{eqn:corf}
into Eq. \ref{eqn:diff1} gives
\begin{equation}
\int_{0}^{t}\int_{0}^{t}\<f(t')f(t")\>dt'dt"=2c_{0}\tau_{d} t.
\end{equation}
Setting this equal to the left-hand side of Eq. \ref{eqn:diff1} yields
\begin{equation}
\gamma^{2}\<(\Delta x)^{2}\> =2c_{0}\tau_{d} t,
\label{eqn:diff2}
\end{equation}
with the diffusion coefficient defined as $\frac{1}{2}\<(\Delta x)^{2}\>$, so that
\begin{equation}
D=\frac{c_{0}\tau_{d}}{\gamma^{2}}.
\label{eqn:diff3}
\end{equation}
Using Eq. \ref{eqn:diff3} and the Einstein Relation, Eq. \ref{eqn:einstein} yields an effective temperature
\begin{equation}
k_{B}T=\frac{c_{0}\tau_{d}}{\gamma}.
\label{eqn:temp}
\end{equation}

\subsection{Estimating microtubule fluctuation amplitude}

To estimate the distance a microtubule moves before encountering a potential barrier of order $k_{B}T$, we set Eq. \ref{eqn:potfluct} equal to $(Ak_{B}T)^{2}$, where $A$ is a multiplicative factor, giving
\begin{equation}
\<V^{2}\>=(Ak_{B}T)^{2}=\frac{2}{3}c\Delta^{2}m^{3}.
\label{eqn:barrier}
\end{equation}
Solving for $m$ yields,
\begin{equation}
m=[\frac{3(Ak_{B}T)^{2}}{2c\Delta^{2}}]^{\frac{1}{3}}.
\end{equation}
Plugging in Eq. \ref{eqn:temp} yields,
\begin{equation}
m=[\frac{3(A\frac{c_{0}\tau_{d}}{\gamma})^{2}}{2c\Delta^{2}}]^{\frac{1}{3}}=[\frac{3(\frac{A}{\gamma}p(1-p)[kf_{k}^{2}+(1-k)f_{n}^{2}]\tau_{d})^{2}}{2p^{2}[kf_{k}^{2}+(1-k)f_{n}^{2}]\Delta^{2}}]^{\frac{1}{3}}.
\label{eqn:barrier2}
\end{equation}

If we make the further simplifications that the two motor species exert the same peak force $f_{e}$,  the concentrations of the species are equal, and the probability of a motor exerting a force is $\frac{1}{2}$, then Eq. \ref{eqn:barrier2} becomes
\begin{equation}
m=[\frac{3(A\frac{f_{e}^{2}\tau_{d}}{4\gamma})^{2}}{\frac{1}{2}f_{e}^{2}\Delta^{2}}]^{\frac{1}{3}}\approx[\frac{(A^{2}f_{e}^{2}\tau_{d}^{2}}{2\gamma^{2}\Delta^{2}}]^{\frac{1}{3}}
\label{eqn:barrier3}
\end{equation}
According to Ref. \cite{Tawada}, $\gamma$ can be approximated as
$kt$, where $k$ is the effective spring constant of the motor, and
t is the characteristic time for the motor to be associated with
the microtubule per cycle \cite{lisup}. Since we approximated
$\tau_{d}\approx t$, Eq. \ref{eqn:barrier3} becomes
\begin{equation}
m\approx[\frac{Af_{e}}{\sqrt{2}k\Delta}]^{\frac{2}{3}}.
\label{eqn:barrier4}
\end{equation}
Using $k\approx1$ pN/nm \cite{lisup}, $\Delta\approx50$ nm \cite{li}, and $f_{e}\approx10$ pN, 
\begin{equation}
m\approx(\frac{A}{2})^{\frac{2}{3}}.
\end{equation}
Therefore, to reach a potential barrier of order $10k_{B}T$, the
microtubule would have to move $\approx 3$ motor sites, or $0.15$
$\mu$m.

This is in agreement with the fluctuation size of $\approx 0.2 \mu m$ measured in
motility assays, as in
Fig. 5 from Ref. \cite{li}. Also, it is likely
that the potential barrier must be greater than  $10k_{B}T$ to
contain the microtubule, thereby giving an estimate closer to the
experimental result.

In addition, we can determine how the distance the microtubule moves
scales with time. According to Kramer's theory of thermal activation,
the time scale $t$ for escaping a potential is  proportional to
$\exp{t\Delta V/T}$ \cite{kramer}. Since $\Delta V\sim x^{3/2}$,
therefore $x\sim \ln{t^{2/3}}$. Thus, the microtubule can cover a
large distance very quickly, but then is trapped in a potential
well and restricted to oscillatory motion.

\begin{figure}[htp]
\begin{center}
(a)\includegraphics[width=0.47\hsize]{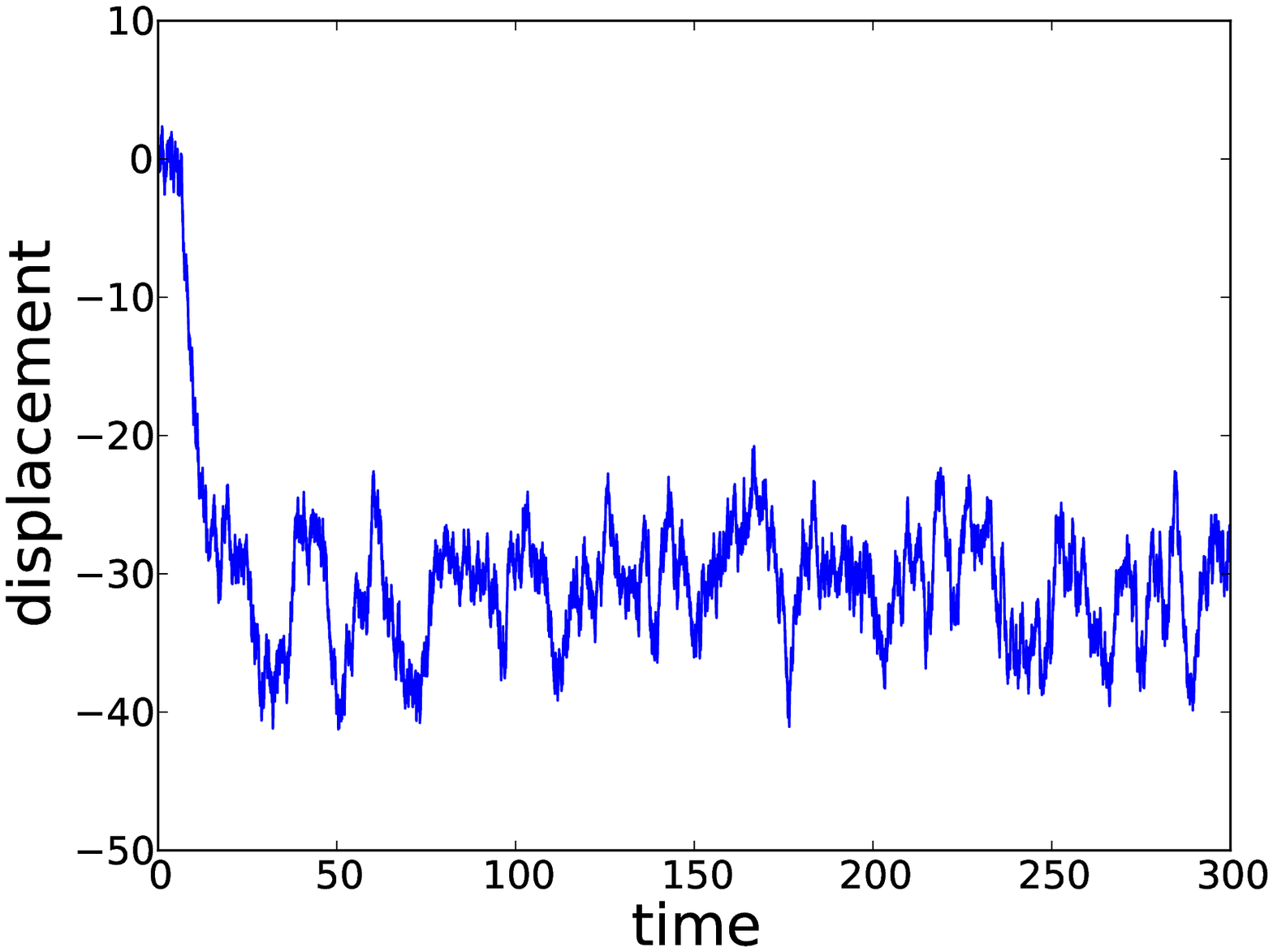}
(b)\includegraphics[width=0.47\hsize]{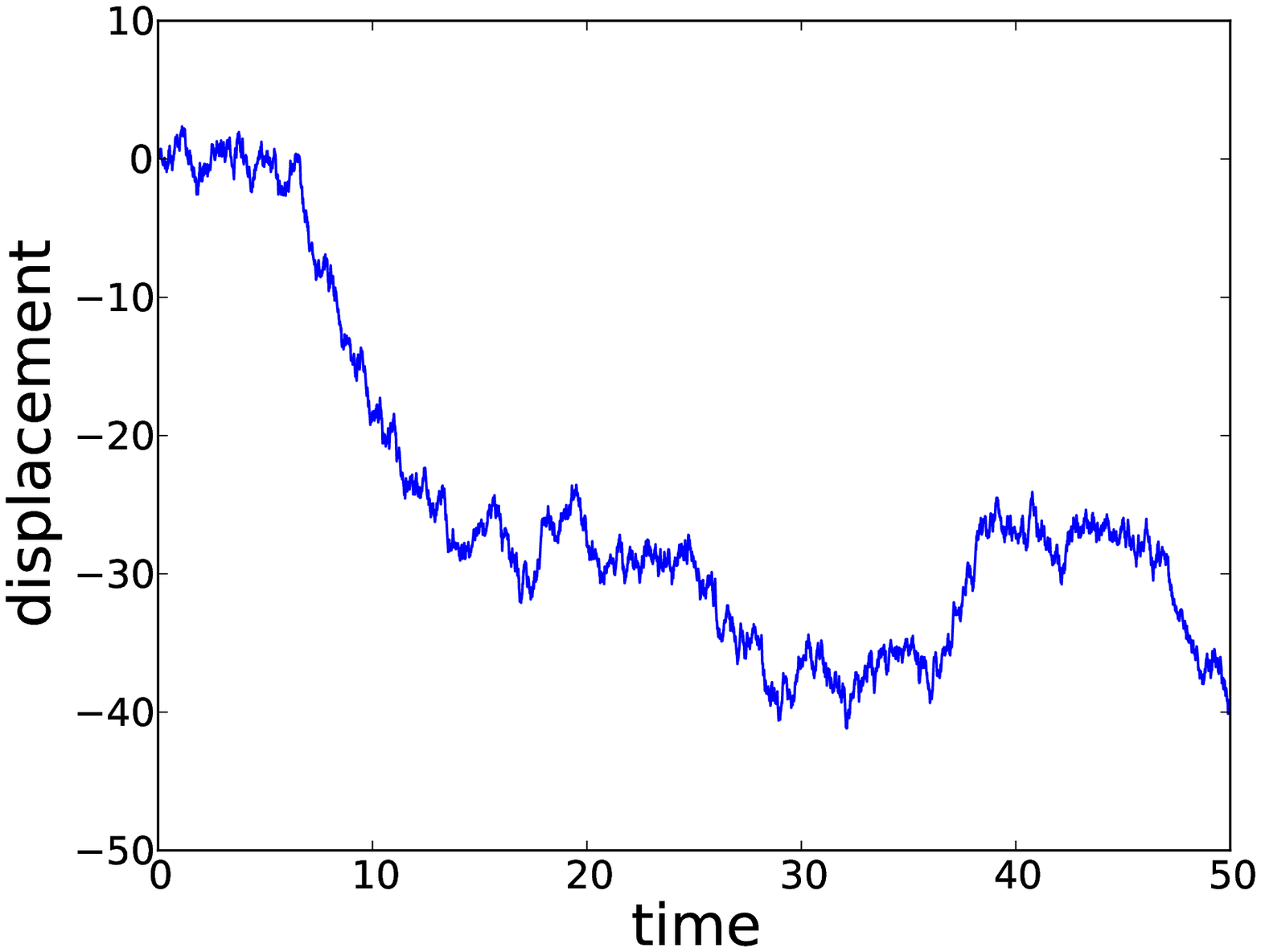}
\caption{
(a) Displacement versus time for a particular realization
of random forces. (b) Same plot but expanded for short times. Note
that initially it moves until it reaches a position which is more favorable
``energetically".
}
\label{fig:displ_vs_t}
\end{center}
\end{figure}

\section{Simulations}

The analytical work of the last section can be taken further by a
numerical implementation of the model described by Eq. \ref{eqn:rwre}.
We use units of length so that the distance between adjacent motors
is unity.

We consider a force produced by a motor at the $ith$ site, $f_i$, drawn from a
standard normal distribution.
The net force acting on a microtubule, $F(j)$, is the sum over $L$ adjacent sites of
these random forces, as in Eq. \ref{eqn:fnet}.
We then linearly interpolate for non-integral values of $x$ to
obtain the force at an arbitrary position which gives us the complete
force for any value of $x$. We choose $n(t)$ from a Gaussian
distribution with standard deviation $C$, to describe the system
at a temperature $T$. We solve Eq. \ref{eqn:rwre} by a simple Euler
discretization with a time step $dt = 0.01$. The noise amplitude
is related to the temperature by $C = \sqrt{2 T/dt}$.

\begin{figure}[htp]
\begin{center}
\includegraphics[width=0.5\hsize]{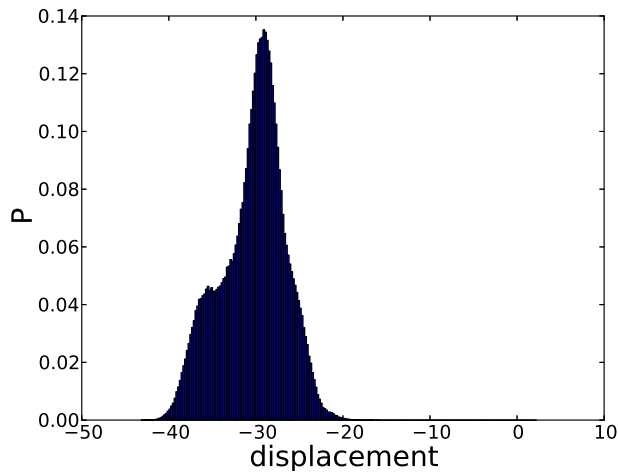}
\caption{
Histogram of positions. The initial transient was not included.
}
\label{fig:displ_hist}
\end{center}
\end{figure}

Fig. \ref{fig:displ_vs_t} shows a single run for one a random realization of
random forces. The displacement starting from $x=0$ is shown as a function of
time in (a). Fig. \ref{fig:displ_vs_t}(b) shows the same data rescaled to 
reveal the behavior of the initial transient. Note that the microtubule moves
approximately 35 units before finding a deep potential minimum.
Figure \ref{fig:displ_hist} displays the corresponding histogram of microtubule
position. The initial transient behavior was not included. The non-Gaussian
shape is due to the underlying roughness of the force $F(x)$ and its
corresponding potential. For extremely long times, this histogram will change
because the microtubule will eventually be able to overcome enormous energy
barriers. However the corresponding times for these are exponentially large
as discussed in the previous section. Under experimentally reasonable time
scales, we expect this kind of histogram to be obtained. 

\begin{figure}[htp]
\begin{center}
\includegraphics[width=0.5\hsize]{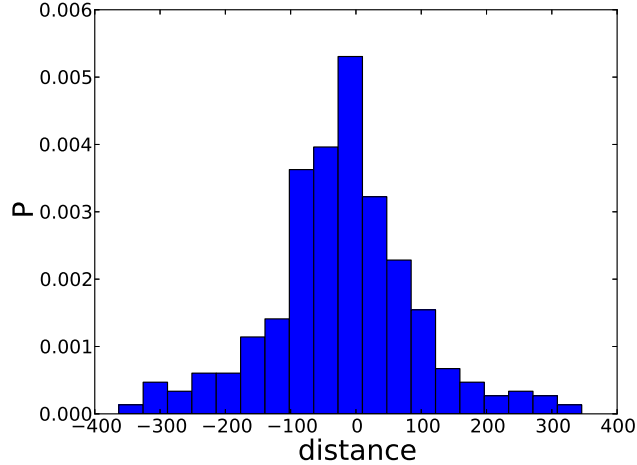}
\caption{
A histogram of the displacement due to the initial transients. The 
simulation was run $300$ times and the initial displacements were all
computed. The initial displacement was obtained by taking the difference
between the steady state average position and the initial position
of the microtubule. 
}
\label{fig:init_shift_hist}
\end{center}
\end{figure}

We now probe the behavior of the initial transients, noted above. We ran this
simulation $300$ times each with randomly generated forces. Then we computed the
initial displacement for each realization by taking the difference between the
steady state average position and the initial position. A histogram of these
differences is shown in Fig.  \ref{fig:init_shift_hist}. As is apparent, the
displacement in a transient is typically about $100$, or the length of the
microtubule. 

\section{Model application to interpolar microtubules during metaphase}

We have seen that the effect of antagonistic motors is indeed to lead to
a quasi-steady state behavior for gliding assays where microtubules fluctuate
in a localized region. However there is a sizeable initial transient
displacement that appears to be of order the length of the microtubule
before it becomes localized.

The reason for this relatively large initial transient can be understood
from the statistics of random walks. The time averaged net force $F$, is the
sum of the forces of the individual motors, Eq. \ref{eqn:fnet}, and the displacement from that
point will follow random walk statistics for lengths smaller than the
length of the microtubule $L$ as seen from Eq. \ref{eqn:pot1}. Typically
$F$ will have a magnitude that scales as $\sqrt{L}$. If we are interested
in how far a microtubule must travel before $F = 0$, then this
must be a distance that is proportional to $L$, because this will typically
give rise to a change in force that is also proportional to  $\sqrt{L}$. 

\begin{figure}[htp]
\begin{center}
\includegraphics[width=0.5\hsize]{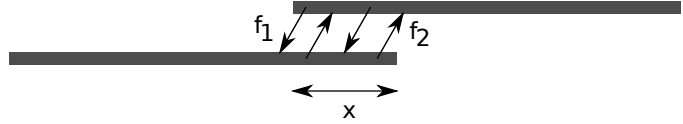}
\caption{
Two anti-parallel microtubules interacting via two types of antagonistic
motors. The motors interact in the region of overlap.
}
\label{fig:apmt}
\end{center}
\end{figure}

Now consider the slightly different situation of competitive motors 
interacting on antiparallel microtubules, interacting with each
other as shown in Fig. \ref{fig:apmt}.
The motors will only operate in the region where the microtubules
overlap, which in the figure is over a distance of length $x$, corresponding
to $n = x/\Delta$ motors.  Eqn.  \ref{eqn:fnet} is modified to be
\begin{equation}
F(m) = \sum_{i=0}^{m}\eta_{i},
\label{eqn:fnetAP}
\end{equation} 
where $\eta_{i} = f_{1,i}+f_{2,i}$ is the sum of the effects motors walking
on both microtubules. This form is different from the gliding assay case, because now the number
of motors included in the sum depends on the overlap distance $x$. However
$F(m)$ is still of the form of a random walk. Therefore the total force is not expected
to vanish except possibly at a finite number of random values of $x$. The
microtubules are therefore expected to have transients that are also large,
but different than the case of the gliding assays.

Because the force has the form of a random walk, and we are interested in where
it passes through zero, this problem is equivalent to the first passage time
problem of a random walk where here the position of the random walk is in force space and time
for the random walks becomes the distance of overlap $x$. In the case where the microtubule move to increasing
$x$, this would be equivalent to a random walk that starts off some distance from zero and asking how
long it will take to cross zero. 
In this case, if the distance of overlap starts out as $x_0$, it will experience a random force $F_0$,
which is the sum of separate motor forces, see  \ref{eqn:fnetAP}.
In the absence of external forces, the microtubules will slide until
they reach an overlap $x^*$ such that $F(x^*) = 0$. As in the gliding assay case, this will be a distance
of order $x_0$. 

On the other hand, if the microtubule moves in a direction of decreasing $x$, then
the statistics of $F(m)$ will be that of a random walk that starts out with $F(0) = 0$.
The probability of a random walk of length $m$ starting at the origin but never passing through
zero can be obtained ~\cite{chandrasekhar}. For example, if the motor spacing
$\Delta = 50 nm$ and the overlap $x= 1 \mu m$, then we are asking for the
probability that a random walk of length $20$ never passes through its starting
position. In that case that probability is about $0.18$. This means that many overlapping
microtubules will be pushed away from each other so that they no longer overlap. 
Many others will increase from an overlap of $1 \mu m$ to greater than $2 \mu
m$. 

\begin{figure}[htp]
\begin{center}
\includegraphics[width=0.5\hsize]{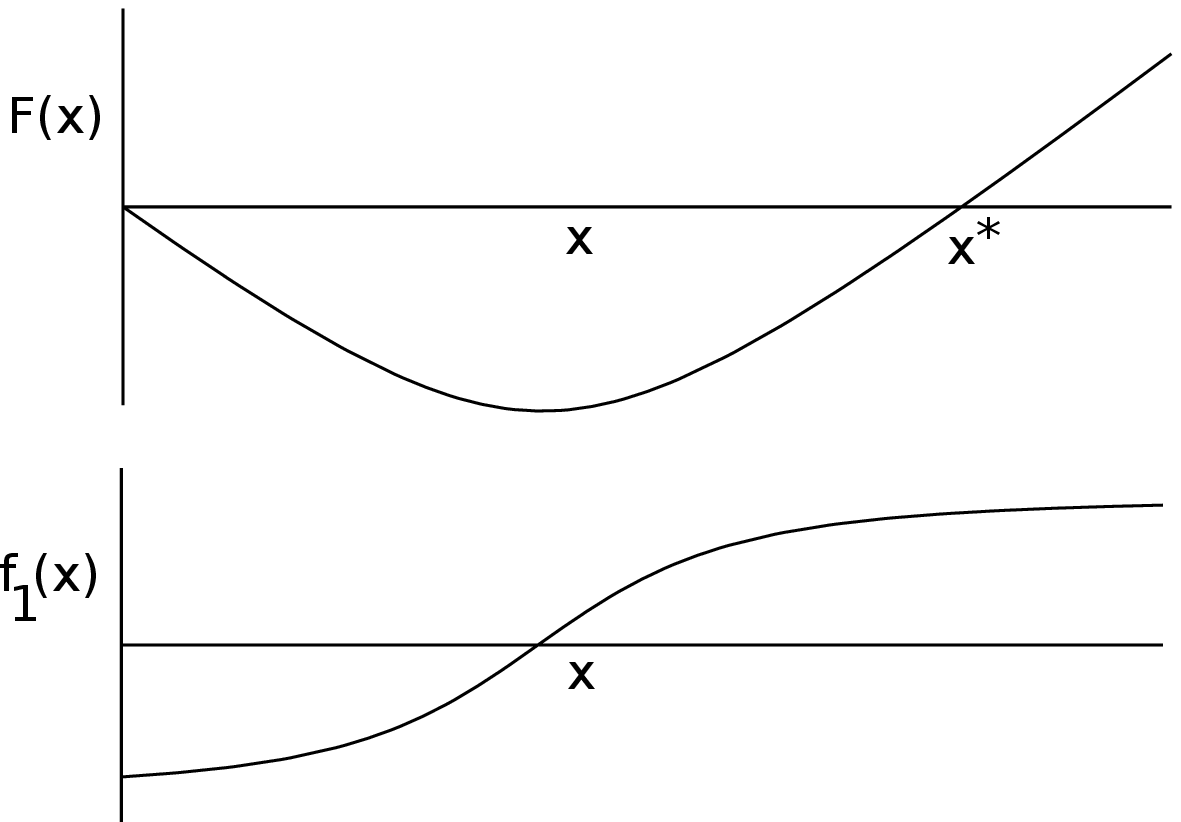}
\caption{
}
\label{fig:force}
\end{center}
\end{figure}

Therefore this model of motor association during prometaphase,
does a somewhat imperfect job of maintaining spindle length.
One simple way to improve its efficacy is to have a {\em gradient} in motor
concentrations. In this case in Eq. \ref{eqn:fnetAP}, instead of
the forces $f_1$ and $f_2$ being random variables with a
mean zero, we say that the local motor concentrations are not
at the balance point so that there is a net time averaged
force that varies deterministically with position. Let us further
assume that the concentration profile is the same for both microtubules.
Then the corresponding force densities $f_1(x)$ and $f_2(x)$, will give
equal contributions to the net force that the microtubules exert on each
other. 
\begin{equation}
F(x) = 2\int_0^x f_1(x') dx' .
\label{eqn:fnetAPx}
\end{equation} 
Fig. \ref{fig:force} displays a force density profile that has the right
properties. $f_1(x)$ starts of negative, close to $x=0$, meaning that near
the tip of the microtubules, the motors  have the net effect of pulling
each other closer together. At some point the sign of $f_1(x)$ changes
and motors in that region predominantly pull the microtubules apart. The
net force due to all the motors as a function of overlap, $F(x)$ is obtained
by integrating $f_1(x)$. The equilibrium overlap $x^*$ is when $F(x^*) = 0$.
By adjusting the profile of motor concentrations, $x^*$ can be shifted.

\subsection{Strength of concentration imbalance}

One caveat that should be mentioned is that if the motor concentration gradients
are too weak, then the microtubules will become stuck due to the randomly
fluctuating force. Without a strong enough bias due to motor concentration
variation, the extremely jagged random
environment will prevent microtubules from sliding to the equilibrium
point $x^*$. We now consider the minimum size of this concentration
variation necessary to overcome the random static forces.

The force in Fig. \ref{fig:force} is linear around the equilibrium point $x^*$,
and is of the form of Hooke's law: $F = -\kappa x$, where $x$ is measured relative
to $x^*$. The maximum value that $\kappa$ can take can be determined by
the extreme case where the force on the microtubule changes from $0$ to $f_e$
within one motor spacing $\Delta$. If this change occurs over a width of $n$ 
motors instead, we can write $\kappa = f_e/(n \Delta)$. We wish to determine
the maximum value of $n$ that is consistent with the microtubules being able
to slide relative to each other. The potential corresponding to Hooke's law is
$U = \kappa x^2/2$. Compared with the statistics of the random 
potential, Eq. \ref{eqn:barrier}, where $V \propto x^{3/2}$, we see
that for small enough $x$, the random force dominates, but when $x$ becomes
sufficiently large, Hooke's law prevails. If the crossover occurs at too large
a value of $x$, then the system will become trapped in some local random minimum
and not be able to move closer to $x^*$. To determine this crossover, we equate
the standard deviation of the fluctuation in the random potential, given by
Eq. \ref{eqn:barrier},  with the Hookean potential.  In terms of $m =x/\Delta$,
we obtain
\begin{equation}
n^2 = \frac{3}{8} \frac{f_e^2 m}{c},
\end{equation}
where the force variance $c$ is given in Eq. \ref{eq:c}. To estimate the value
of $n$, we take $p=1/2$, and $f_k=f_n=f_e$. This gives $n^2 = (3/2) m$. 
Experimentally in a gliding assay, it was found that the fluctuations in $m$ where approximately 4. 
This means that we expect that $n$ should be in the range 2 to 3, in order for
the microtubules to slide within the experimental time scale. 

This small value of $n$ is consistent with the fact that at the balance point,
the fluctuations in position of a microtubule in a gliding assay are small. 
The spacing between motors, in these assays, is presumably not identical to interpolar
microtubules in the spindle but they are believed to be plausibly similar~\cite{li,BrustMascher}.

\section{Conclusion}

By means of a fairly general analysis, making few assumptions, we
have seen that the motion of microtubules in an antagonistic motility
assay near the balance point, is well described by a random walk in
a random environment,\cite{rwre} but with large correlations in the static random
force. The connection does not depend on a detailed model of the motors. The
force velocity curves of the motors near the stall point, $v=0$, only
come into play to give a drag coefficient for the microtubule. The results
found are in reasonable agreement with the fluctuations observed in
microtubule positions seen in experiments~\cite{li}. However the
model also predicts that the microtubule will slide of order their
own distance before getting badly trapped.

There is substantial similarity between these gliding assay experiments
and models for what occurs {\em in vivo} during prometaphase, with
KLP61F and Ncd motors antagonizing each other in mitotic
spindles~\cite{li,civelekoglu}.  However we point out that although their
model correcly predicts that these motors will act to inhibit variations
in spindle length, they also do so imperfectly. Antiparallel
interpolar microtubules will have initial transients where they
slide of order their own length, or sometimes completely disassociate.
It appears that such variations are within the error bars of
experimental observation however.
But, to circumvent these fluctuations, it is possible that the cell
sets up a gradient of motors during prometaphase, thereby restricting
the mitotic spindle to oscillatory movements in a potential well.
Further observations of the densities of motors {\em in vivo} could
lend a great deal to further understanding of this situation.

\section{Acknowledgments}
This material is based upon work supported by
the National Science Foundation under Grant CCLI DUE-0942207.

\end{document}